\documentclass{article}

\newcommand{\ba}{\begin{eqnarray}}
\newcommand{\ea}{\end{eqnarray}}
\begin{document}
\pagestyle{plain}

\title{Masses and magnetic moments of pentaquarks}
\author{R. Bijker\\
Instituto de Ciencias Nucleares, 
Universidad Nacional Aut\'onoma de M\'exico,\\
A.P. 70-543, 04510 M\'exico, D.F., M\'exico
\and
M.M. Giannini and E. Santopinto \\
Dipartimento di Fisica dell'Universit\`a di Genova, 
I.N.F.N., Sezione di Genova, \\
via Dodecaneso 33, 16164 Genova, Italy}
\date{}
\maketitle                   

\begin{abstract}
We discuss the spectroscopy of $q^4 \bar{q}$ pentaquarks. The quantum 
numbers of the ground state depend on the interplay between spin-flavor 
and orbital contributions to the energy. The magnetic moments of the 
lowest pentaquark state with negative and positive parity are found to 
be 0.382 $\mu_N$ and 0.089 $\mu_N$, respectively. 
\end{abstract}

\begin{center}
\end{center}

\section{Introduction}

The discovery of the $\Theta^+(1540)$ resonance with positive strangeness 
$S=+1$ by the LEPS Collaboration \cite{leps} and the subsequent 
confirmation by various other experimental collaborations 
\cite{diana,saphir,clas,itep,hermes} has sparked an enormous amount of 
experimental and theoretical studies of exotic baryons \cite{jlab}. 
Since there are no three-quark configurations with these properties, it 
is an exotic baryon resonance, whose simplest configuration is that of  
a $q^4 \bar{q}$ pentaquark consisting of four quarks and an antiquark. 
Its other quantum numbers are still mostly unknown, although the absence of 
a signal for a $\Theta^{++}$ pentaquark state in the $pK^+$ invariant mass 
spectrum is an indication that the observed $\Theta^+$ is most 
likely to be an isosinglet $I=0$ \cite{saphir,clas,hermes}. 
The width of the $\Theta^+$ is so small that only an upper limit could be  
established. A re-analysis of elastic $K^+N$ scattering data has put an even 
more stringent upper bound on the width of only a few MeV \cite{width} 
which suggests that it is exotic dynamically 
as well as in its quantum numbers. 
More recently, evidence has been found for the existence of another exotic 
baryon $\Xi^{--}(1862)$ with strangeness $S=-2$ by the NA49 Collaboration 
\cite{cern}. The $\Theta^+$ and $\Xi^{--}$ resonances are 
interpreted as pentaquarks belonging to a flavor antidecuplet with 
isospin $I=0$ and $I=3/2$, respectively. 

The spin and parity of the $\Theta^+$ have not yet been determined 
experimentally. Theoretically, there seems to be agreement between different 
approaches on the value of the angular momentum, i.e. $J=1/2$, but not so on 
the parity. The parity of the 
lowest pentaquark state is predicted to be positive by many studies, such as 
chiral soliton models \cite{soliton}, cluster models \cite{JW,SZ,KL} and various 
constituent quark models \cite{cqmpos}. However, there is also evidence for 
a negative parity ground state pentaquark from recent work on QCD sum rules 
\cite{sumrule}, lattice QCD \cite{lattice} and a quark model calculation 
\cite{cqmneg}. Various suggestions have been made for future experiments 
to measure the parity \cite{parity,Nam,Zhao,photo}.

Another unknown quantity is the magnetic moment. Although it may be difficult 
to determine its value experimentally, it is an essential ingredient in 
calculations of the photo- and electroproduction cross sections 
\cite{Nam,Zhao,photo}. In the absence of experimental information, 
one has to rely on model calculations. 

The aim of this article is to study the spin, parity and magnetic moment 
of the ground state pentaquark. First, we construct a complete classification 
scheme of pentaquark states based on the spin-flavor $SU(6)$ symmetry. 
Some general characteristics of the spectrum are discussed for a schematic 
harmonic oscillator quark model. Next, we derive the magnetic 
moments of the lowest exotic pentaquarks with negative and positive parity. 
Finally, it is shown that in a string-like algebraic model the ground state 
has negative parity. 

\section{Pentaquark states}

We consider pentaquarks to be built of five constituent parts which are 
characterized by both internal and spatial degrees of freedom 
(see Fig.~\ref{gsbar}). 

\subsection{Internal degrees of freedom}

The internal degrees of freedom are taken to be the three light flavors 
$u$, $d$, $s$ with spin $s=1/2$ and three possible colors $r$, $g$, $b$.   
The corresponding algebraic structure consists of the usual spin-flavor 
and color algebras
\ba
{\cal G}_i \;=\; SU_{\rm sf}(6) \otimes SU_{\rm c}(3) ~.
\ea
The pentaquark spin-flavor states can be classified according to the 
irreducible representations of $SU_{\rm sf}(6)$ and its subgroups 
\ba 
\left| \begin{array}{ccccccccccc}
SU_{\rm sf}(6) &\supset& SU_{\rm f}(3) &\otimes& SU_{\rm s}(2) 
&\supset& SU_{\rm I}(2) &\otimes& U_{\rm Y}(1) &\otimes& SU_{\rm s}(2) \\
\, [f] &,& [g] &,& [h] &,& I &,& Y && \end{array} \right> ~,  
\label{sfchain}
\ea
where $I$ and $Y$ denote the isospin and hypercharge. 

\begin{figure}
\centering
\setlength{\unitlength}{0.8pt}
\begin{picture}(350,200)(75,75)
\thinlines
\put( 75,150) {\vector(1,0){350}}
\put(250, 75) {\vector(0,1){225}}
\put(100,145) {\line(0,1){10}}
\put(150,145) {\line(0,1){10}}
\put(200,145) {\line(0,1){10}}
\put(250,145) {\line(0,1){10}}
\put(300,145) {\line(0,1){10}}
\put(350,145) {\line(0,1){10}}
\put(400,145) {\line(0,1){10}}
\thicklines
\put(200,200) {\line(1,0){100}}
\put(150,150) {\line(1,0){200}}
\put(100,100) {\line(1,0){300}}

\put(100,100) {\line(1,1){150}}
\put(200,100) {\line(1,1){100}}
\put(300,100) {\line(1,1){ 50}}

\put(200,100) {\line(-1,1){ 50}}
\put(300,100) {\line(-1,1){100}}
\put(400,100) {\line(-1,1){150}}

\multiput(250,250)(100,0){1}{\circle*{5}}
\multiput(200,200)(100,0){2}{\circle*{5}}
\multiput(150,150)(100,0){3}{\circle*{5}}
\multiput(100,100)(100,0){4}{\circle*{5}}

\multiput(100,100)(300,0){2}{\circle*{10}}
\put(250,250){\circle*{10}}

\put(260,250){$\Theta^+: \; uudd\bar{s}$}
\put( 70, 75){$\Xi_{3/2}^{--}: \; ddss\bar{u}$}
\put(375, 75){$\Xi_{3/2}^+: \; uuss\bar{d}$}

\put(425,125){$I_3$}
\put(225,300){$Y$}
\put(100,165){\makebox(0,0){$-\frac{3}{2}$}}
\put(150,165){\makebox(0,0){$-1$}}
\put(200,165){\makebox(0,0){$-\frac{1}{2}$}}
\put(300,165){\makebox(0,0){$ \frac{1}{2}$}}
\put(350,165){\makebox(0,0){$ 1$}}
\put(400,165){\makebox(0,0){$ \frac{3}{2}$}}
\put(245,255){\makebox(0,0)[br]{$2$}}
\put(245,205){\makebox(0,0)[br]{$1$}}
\put(245,105){\makebox(0,0)[br]{$-1$}}
\end{picture}
\caption[]{$SU(3)$ flavor multiplet $[33]$ with $E$ symmetry. 
The isospin-hypercharge multiplets are $(I,Y)=(0,2)$, $(\frac{1}{2},1)$, 
$(1,0)$ and $(\frac{3}{2},-1)$. Exotic states are indicated with $\bullet$.}
\label{flavor33e}
\end{figure}
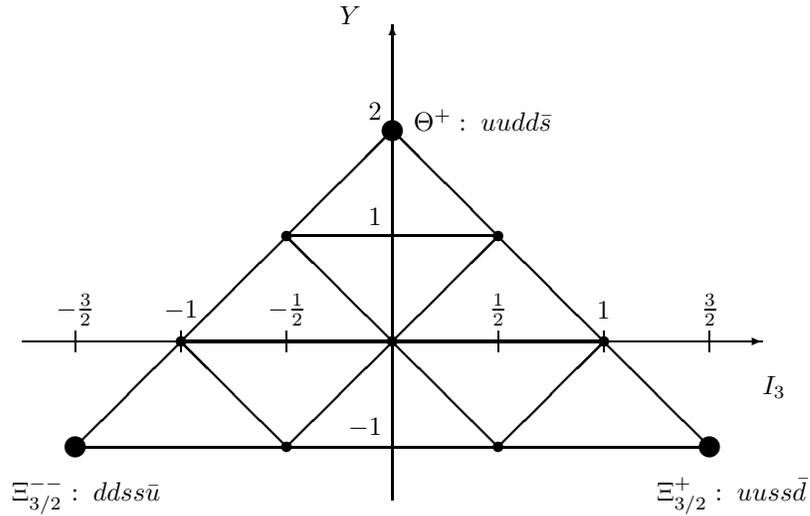

In the construction of the classification scheme we are guided by two 
conditions: the pentaquark wave function should be a color singlet, and 
should be antisymmetric under any permutation of the four quarks. 
The permutation symmetry of the four-quark subsystem is characterized by 
the $S_4$ Young tableaux $[4]$, $[31]$, $[22]$, $[211]$ and $[1111]$ or, 
equivalently, by the irreducible representations of the tetrahedral group 
${\cal T}_d$ (which is isomorphic to $S_4$) as $A_1$, $F_2$, $E$, $F_1$ 
and $A_2$, respectively. For notational purposes we use the latter to 
label the discrete symmetry of the pentaquark wave functions. 
The corresponding dimensions are 1, 3, 2, 3 and 1. 
The allowed spin, flavor and spin-flavor states are obtained by 
standard group theoretic techniques (see Table~\ref{states}).
The full decomposition of the spin-flavor states into spin and flavor states 
is given in Table~5 of \cite{BGS}. The states of a given flavor multiplet 
can be labeled by isospin $I$, $I_3$ and hypercharge $Y$. 
The electric charge is given by the Gell-Mann-Nishijima relation 
\ba
Q \;=\; I_3 + \frac{Y}{2} \;=\; I_3 + \frac{B+S}{2} ~, 
\ea 
where $B$ denotes the baryon number and $S$ the strangeness.   
It is difficult to distinguish the pentaquark flavor singlets, octets 
and decuplets from the three-quark flavor multiplets, since they have 
the same values of the hypercharge $Y$ and isospin projection $I_3$. The same 
observation holds for the majority of the states in the remaining flavor 
states. However, the antidecuplets, the 27-plets and 35-plets contain 
in addition exotic states which cannot be obtained from three-quark 
configurations. These states are more easily identified experimentally 
due to the uniqueness of their quantum numbers. As an example, the exotic 
states of the antidecuplet are indicated by a $\bullet$ in 
Fig.~\ref{flavor33e}: the $\Theta^+$ is the isosinglet $I=I_3=0$ with 
hypercharge $Y=2$ (strangeness $S=+1$), and the cascades   
$\Xi_{3/2}^+$ and $\Xi_{3/2}^{--}$ 
have hypercharge $Y=-1$ (strangeness $S=-2$) and isospin $I=3/2$ with 
projection $I_3=3/2$ and $-3/2$, respectively. 

\subsection{Spatial degrees of freedom}

The relevant degrees of freedom for the relative motion of the 
constituent parts are provided by the Jacobi coordinates 
which we choose as \cite{KM}
\ba
\vec{\rho}_1 &=& \frac{1}{\sqrt{2}} 
( \vec{r}_1 - \vec{r}_2 ) ~,
\nonumber\\
\vec{\rho}_2 &=& \frac{1}{\sqrt{6}} 
( \vec{r}_1 + \vec{r}_2 - 2\vec{r}_3 ) ~,
\nonumber\\
\vec{\rho}_3 &=& \frac{1}{\sqrt{12}} 
( \vec{r}_1 + \vec{r}_2 + \vec{r}_3 - 3\vec{r}_4 ) ~,
\nonumber\\
\vec{\rho}_4 &=& \frac{1}{\sqrt{20}} 
( \vec{r}_1 + \vec{r}_2 + \vec{r}_3 + \vec{r}_4 - 4\vec{r}_5 ) ~, 
\label{jacobi}
\ea
where $\vec{r}_i$ ($i=1,..,4$) denote the coordinate of the $i$-th quark, 
and $\vec{r}_5$ that of the antiquark. The last Jacobi coordinate 
is symmetric under the interchange of the quark coordinates, 
and hence transforms as $A_1$ under ${\cal T}_d$ ($\sim S_4$), whereas 
the first three transform as three components of $F_2$ \cite{KM}. 

\subsection{Pentaquark wave functions}

The pentaquark wave function is obtained by combining the spin-flavor part 
with the color and orbital parts in such a way that the total wave function 
is a color-singlet state, and that the four quarks satisfy the Pauli 
principle, i.e. are antisymmetric under any permutation of the four quarks. 
Since the color part of the pentaquark wave function is a $[222]$ singlet 
and that of the antiquark a $[11]$ anti-triplet, the color wave function of 
the four-quark configuration is a $[211]$ triplet which has $F_1$ symmetry 
under ${\cal T}_d$. The total $q^4$ wave function is antisymmetric ($A_2$), 
hence the orbital-spin-flavor part has to have $F_2$ symmetry
\ba
\psi_{A_2} \;=\; \left[ \psi^{\rm c}_{F_1} \times 
\psi^{\rm osf}_{F_2} \right]_{A_2} ~.
\label{wf}
\ea
Here the square brackets $[\cdots]$ denote the tensor coupling under the 
tetrahedral group ${\cal T}_d$. 
In Table~\ref{tdpenta}, we present the allowed spin-flavor multiplets with 
exotic pentaquarks for some lowlying orbital excitations. The exotic 
spin-flavor states associated with the $S$-wave state $L^p_t=0^+_{A_1}$ 
all belong to the $[f]_t=[42111]_{F_2}$ spin-flavor multiplet.
The corresponding orbital-spin-flavor wave function is given by 
\ba
\psi^{\rm osf}_{F_2} \;=\;
\left[ \psi^{\rm o}_{A_1} \times \psi^{\rm sf}_{F_2} \right]_{F_2} ~.
\label{wf1} 
\ea
A $P$-wave radial excitation with $L^p_t=1^-_{F_2}$ gives rise to exotic 
pentaquark states of the $[51111]_{A_1}$, $[42111]_{F_2}$, $[33111]_E$ and 
$[32211]_{F_1}$ spin-flavor configurations. They are characterized by 
the orbital-spin-flavor wave functions 
\ba
\psi^{\rm osf}_{F_2} \;=\;
\left[ \psi^{\rm o}_{F_2} \times \psi^{\rm sf}_{t} \right]_{F_2} ~,
\label{wf2}
\ea
with $t=A_1$, $F_2$, $E$ and $F_1$, respectively.

\section{Harmonic oscillator quark model}

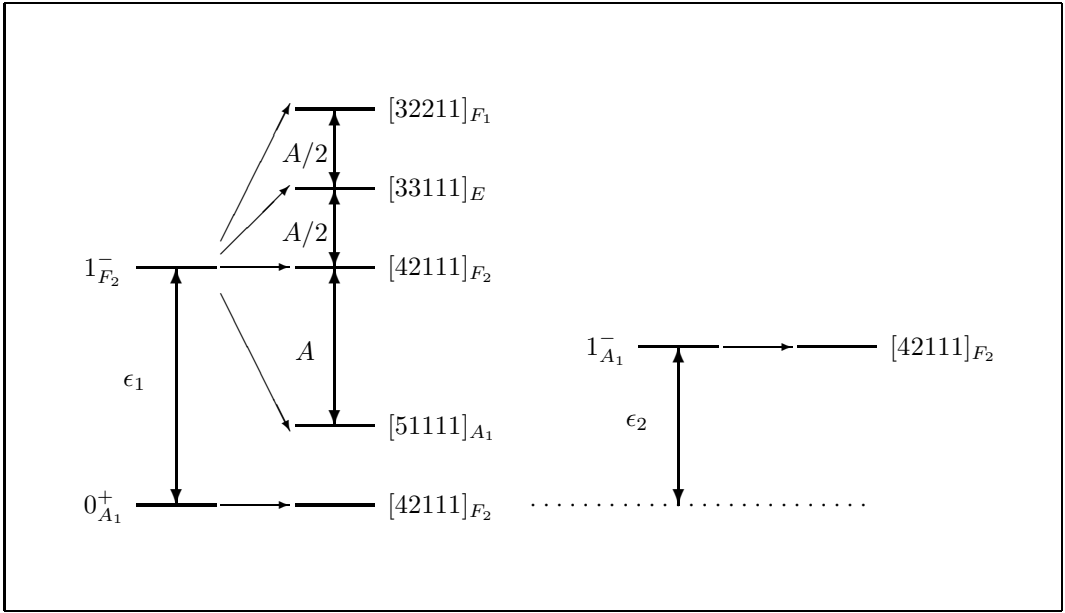
\begin{figure}
\centering
\setlength{\unitlength}{1.0pt}
\begin{picture}(400,230)(0,30)
\thinlines
\put(  0, 30) {\line(1,0){400}}
\put(  0,260) {\line(1,0){400}}
\put(  0, 30) {\line(0,1){230}}
\put(400, 30) {\line(0,1){230}}
\put( 82, 70){\vector(1, 0){26}}
\put(272,130){\vector(1, 0){26}}
\put( 82,170){\vector(1, 2){26}}
\put( 82,165){\vector(1, 1){26}}
\put( 82,160){\vector(1, 0){26}}
\put( 82,150){\vector(1,-2){26}}

\thicklines
\put( 50, 70) {\line(1,0){ 30}}
\put(110, 70) {\line(1,0){ 30}} 
\put( 50,160) {\line(1,0){ 30}}
\put(110,100) {\line(1,0){ 30}}
\put(110,160) {\line(1,0){ 30}}
\put(110,190) {\line(1,0){ 30}}
\put(110,220) {\line(1,0){ 30}}
\put(240,130) {\line(1,0){ 30}}
\put(300,130) {\line(1,0){ 30}}
\put( 65, 70) {\vector(0, 1){90}}
\put( 65,160) {\vector(0,-1){90}}
\put( 45,115) {$\epsilon_1$}
\multiput(200, 70)(5, 0){26}{\circle*{1}}
\put(255, 70) {\vector(0, 1){60}}
\put(255,130) {\vector(0,-1){60}}
\put(235,100) {$\epsilon_2$}
\put(125,100) {\vector(0, 1){60}}
\put(125,160) {\vector(0, 1){30}}
\put(125,190) {\vector(0, 1){30}}
\put(125,160) {\vector(0,-1){60}}
\put(125,190) {\vector(0,-1){30}}
\put(125,220) {\vector(0,-1){30}}
\put(110,125) {$A$}
\put(105,170) {$A/2$}
\put(105,200) {$A/2$}
\put( 30, 67) {$0^+_{A_1}$}
\put(145, 67) {$[42111]_{F_2}$}
\put( 30,157) {$1^-_{F_2}$}
\put(145, 97) {$[51111]_{A_1}$}
\put(145,157) {$[42111]_{F_2}$}
\put(145,187) {$[33111]_{E}$}
\put(145,217) {$[32211]_{F_1}$}
\put(220,127) {$1^-_{A_1}$}
\put(335,127) {$[42111]_{F_2}$}
\end{picture}
\vspace{15pt}
\caption[]{Schematic spectrum of exotic pentaquarks for orbital excitations 
up to $N=1$ quantum calculated using Eq.~(\ref{ener}) with $A < \epsilon_1$ 
and $B=C=D=E=0$. The orbital excitations are labeled by angular momentum 
and parity $L^p_t$ and the spin-flavor multiplets by $[f]_t$, where the 
index $t$ gives the permutational symmetry of the four-quark system. 
The ground state is associated with the $L^p_t=0^+_{A_1}$ orbital state. 
The $1^-_{F_2}$ state arises from the relative motion of the four 
quarks, whereas the $1^-_{A_1}$ state comes from the relative motion 
between the antiquark and the four-quark system.}
\label{orbex1}
\end{figure}

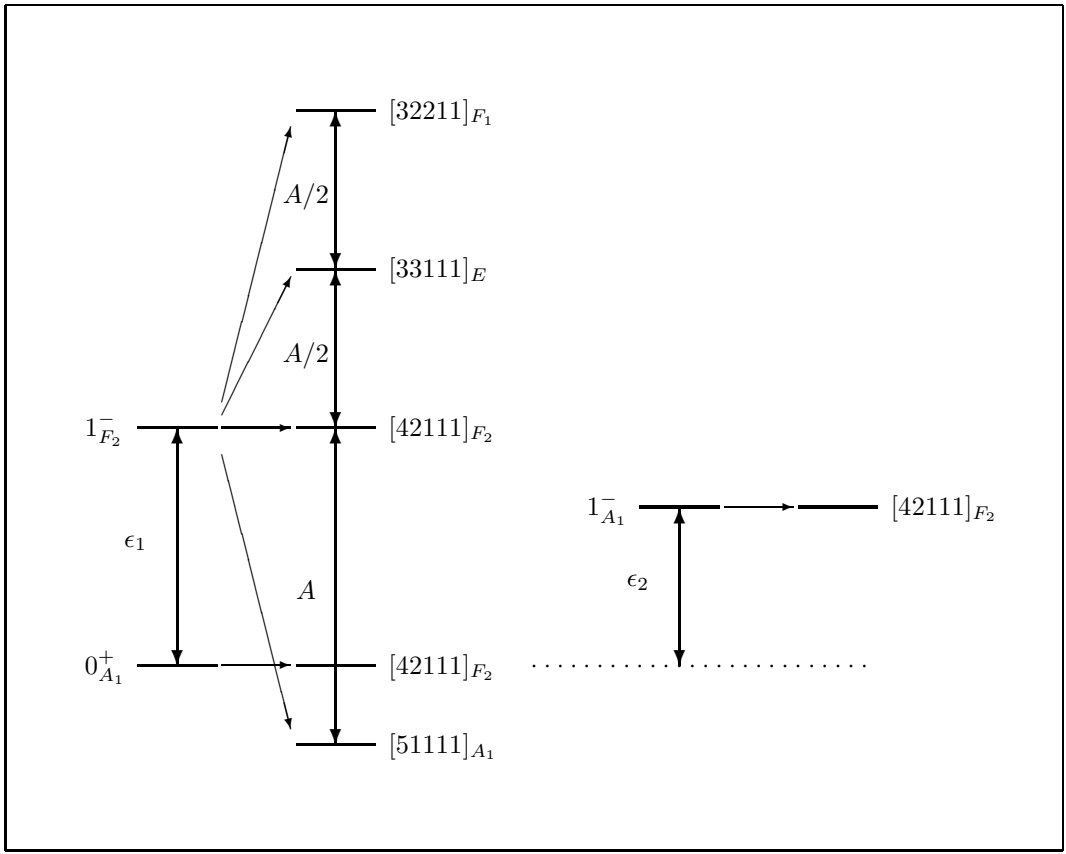
\begin{figure}
\centering
\setlength{\unitlength}{1.0pt}
\begin{picture}(400,320)(0,0)
\thinlines
\put(  0,  0) {\line(1,0){400}}
\put(  0,320) {\line(1,0){400}}
\put(  0,  0) {\line(0,1){320}}
\put(400,  0) {\line(0,1){320}}
\put( 82, 70){\vector(1, 0){26}}
\put(272,130){\vector(1, 0){26}}
\put( 82,170){\vector(1, 4){26}}
\put( 82,165){\vector(1, 2){26}}
\put( 82,160){\vector(1, 0){26}}
\put( 82,150){\vector(1,-4){26}}

\thicklines
\put( 50, 70) {\line(1,0){ 30}}
\put(110, 70) {\line(1,0){ 30}} 
\put( 50,160) {\line(1,0){ 30}}
\put(110, 40) {\line(1,0){ 30}}
\put(110,160) {\line(1,0){ 30}}
\put(110,220) {\line(1,0){ 30}}
\put(110,280) {\line(1,0){ 30}}
\put(240,130) {\line(1,0){ 30}}
\put(300,130) {\line(1,0){ 30}}
\put( 65, 70) {\vector(0, 1){ 90}}
\put( 65,160) {\vector(0,-1){ 90}}
\put( 45,115) {$\epsilon_1$}
\multiput(200, 70)(5, 0){26}{\circle*{1}}
\put(255, 70) {\vector(0, 1){60}}
\put(255,130) {\vector(0,-1){60}}
\put(235,100) {$\epsilon_2$}
\put(125, 40) {\vector(0, 1){120}}
\put(125,160) {\vector(0, 1){60}}
\put(125,220) {\vector(0, 1){60}}
\put(125,160) {\vector(0,-1){120}}
\put(125,220) {\vector(0,-1){60}}
\put(125,280) {\vector(0,-1){60}}
\put(110, 95) {$A$}
\put(105,185) {$A/2$}
\put(105,245) {$A/2$}
\put( 30, 67) {$0^+_{A_1}$}
\put(145, 67) {$[42111]_{F_2}$}
\put( 30,157) {$1^-_{F_2}$}
\put(145, 37) {$[51111]_{A_1}$}
\put(145,157) {$[42111]_{F_2}$}
\put(145,217) {$[33111]_{E}$}
\put(145,277) {$[32211]_{F_1}$}
\put(220,127) {$1^-_{A_1}$}
\put(335,127) {$[42111]_{F_2}$}
\end{picture}
\vspace{15pt}
\caption[]{As Figure~\ref{orbex1}, but for $A > \epsilon_1$. 
The ground state is associated with the $L^p_t=1^-_{F_2}$ state.}
\label{orbex2}
\end{figure}

In a harmonic oscillator treatment, the group structure related to 
the four relative coordinates of Eq.~(\ref{jacobi}) is 
\ba
{\cal G}_r \;=\; U(12) \;\supset\; U(9) \;\otimes\; U(3) ~,
\ea
where $U(9)$ describes the relative motion of the four quarks and 
$U(3)$ that of the antiquark with respect to the four-quark subsystem.  
We consider a simple schematic model for the spectrum of exotic 
pentaquarks in which the Hamiltonian is given by 
\ba
H \;=\; H_{\rm orb} + H_{\rm sf} ~. 
\label{ham}
\ea
Here $H_{\rm orb}$ describes the orbital motion of the 
pentaquark in terms of harmonic oscillators 
\ba
H_{\rm orb} \;=\; \frac{1}{2} \epsilon_1 \sum_{i=1}^3 
\left( \vec{p}_{\rho_i}^{\, 2} + \vec{\rho}_i^{\, 2} \right) 
+ \frac{1}{2} \epsilon_2 \left( \vec{p}_{\rho_4}^{\, 2} 
+ \vec{\rho}_4^{\, 2} \right) ~. 
\label{horb}
\ea
The first term comes from the three degenerate three-dimensional 
harmonic oscillators to describe the relative motion of the four 
quarks, and the second one from the three-dimensional harmonic 
oscillator for the relative motion of the antiquark with respect 
to the four-quark system. The energy eigenvalues of $H_{\rm orb}$ are 
\ba
E_{\rm orb} \;=\; \epsilon_1 \, \left( n_1 + \frac{9}{2} \right) 
+ \epsilon_2 \, \left( n_2 + \frac{3}{2} \right) ~. 
\label{eorb}
\ea
The ground state is an $S$-wave state with $L^p=0^+$ and $A_1$ 
symmetry for the four quarks. Since the orbital excitations are 
described by four relative coordinates, there are four excited 
$P$-wave states with $L^p=1^-$, three of which correspond to 
excitations in the relative coordinates of the four quarks, and 
the fourth to an excitation in the relative coordinate of the 
four-quark system and the antiquark. As a consequence of the 
discrete symmetry of the four quarks, the first three excitations 
form a degenerate triplet with three-fold $F_2$ symmetry, and the 
fourth has $A_1$ symmetry (see the left-hand side of 
Figs.~\ref{orbex1} and~\ref{orbex2}). 
The second term in Eq.~(\ref{ham}) represents the spin-flavor 
dependence of the masses in terms of a generalized 
G\"ursey-Radicati form 
\ba
H_{\rm sf} &=& -A \, C_{2SU_{\rm sf}(6)} 
+ B \, C_{2SU_{\rm f}(3)} + C \, C_{2SU_{\rm s}(2)} 
\nonumber\\
&& + D \, C_{1U_{\rm Y}(1)} 
+ E \, [C_{2SU_{\rm I}(2)}-\frac{1}{4} C_{1U_{\rm Y}(1)}^2] ~, 
\label{hsf}
\ea
with eigenvalues
\ba
E_{\rm sf} &=& - \frac{1}{2} A \left[ \sum_{i=1}^6 f_i(f_i+7-2i) 
- \frac{1}{6} \left(\sum_{i=1}^6 f_i\right)^2 \right] 
\nonumber\\
&& + \frac{1}{2} B \left[ \sum_{i=1}^3 g_i(g_i+4-2i) 
- \frac{1}{3} \left(\sum_{i=1}^3 g_i\right)^2 \right] 
\nonumber\\
&& + C \, s(s+1) + D \, Y + E \, [I(I+1)-\frac{1}{4}Y^2] ~. 
\label{ener}
\ea
The contribution from the flavor part is usually expressed in 
terms of $p=g_1-g_2$ and $q=g_2-g_3$ to obtain 
\ba
\frac{1}{3} B (p^2 + q^2 + 3(p+q) + pq ) ~.
\ea
For the definition of the Casimir operators in Eq.~(\ref{hsf}), we have 
followed the same convention as by Helminen and Riska in \cite{cqmpos}. 
The last two terms correspond to the Gell-Mann-Okubo mass formula 
that describes the splitting within a flavor multiplet \cite{GMO}. 
This formula was extended by G\"ursey and Radicati \cite{GR} to include 
the terms proportional to $B$ and $C$ that depend on the spin and the 
flavor representations, which in turn was generalized further to 
include the spin-flavor term proportional to $A$ as well \cite{BIL}. 

The energy of a given spin-flavor multiplet depends on the oscillator 
frequencies $\epsilon_1$ and $\epsilon_2$, and the coefficient $A$, while 
the terms proportional to $B$, $C$, $D$ and $E$ give the splitting inside 
the multiplet. The sign of the coefficient 
$A$ is taken to be positive corresponding to an attractive 
spin-flavor hyperfine interaction \cite{Riska}, in agreement with 
the sign used in previous studies of baryons as $qqq$ configurations 
\cite{BIL}. The ground state configuration depends on the relative 
size of $\epsilon_1$ and $A$. Its parity is opposite to that of the 
orbital excitation due to the negative intrinsic parity of 
$q^4 \bar{q}$ configurations. For $A < \epsilon_1$, the ground 
state pentaquark is associated with the orbital state $(n_1,n_2)=(0,0)$, 
$L^p_t=0^+_{A_1}$ and the spin-flavor multiplet $[42111]_{F_2}$,  
and therefore has negative parity (see Fig.~\ref{orbex1}). 
For $A > \epsilon_1$, the parity of the lowest pentaquark state 
is positive, since the ground state now corresponds to the orbital 
excitation $(n_1,n_2)=(1,0)$, $L^p_t=1^-_{F_2}$ 
and the spin-flavor state $[51111]_{A_1}$ (see Fig.~\ref{orbex2}). 

The terms proportional to $B$, $C$, $D$ and $E$ only contribute to 
the splitting within a spin-flavor multiplet. They can be estimated  
from mass differences between baryon resonances as  
\ba
3B+3C+3E &=&  \left[ M(\Delta(1232)) - M(N(938)) \right] ~, 
\nonumber\\
C &=& \frac{1}{3} \left[ M(N(1650)) - M(N(1535)) \right] ~, 
\nonumber\\
D &=& \frac{1}{4} \left[ 4M(N(938)) - M(\Sigma(1193)) - 3M(\Lambda(1116)) 
\right] ~,
\nonumber\\
E &=& \frac{1}{2} \left[ M(\Sigma(1193)) - M(\Lambda(1116)) \right] ~. 
\ea
This gives the numerical values $B= 21.3$ MeV, $C=38.3$ MeV, 
$D=-197.3$ MeV and $E= 38.5$ MeV. For the ground state spin-flavor 
configurations $[51111]_{A_1}$ and $[42111]_{F_2}$ in Figs.~\ref{orbex1} 
and~\ref{orbex2}, respectively, the lowest exotic pentaquark state has 
spin $s=1/2$ and belongs to the antidecuplet \cite{BGS}.  

In summary, the parity of the ground state exotic pentaquark 
depends on the relative contribution of the orbital and spin-flavor 
parts of the mass operator. We find that if the splitting due to the 
$SU_{\rm sf}(6)$ spin-flavor term is large compared to that between the 
orbital states, the ground state pentaquark has positive parity, 
whereas for a relatively small spin-flavor splitting the parity of the 
lowest pentaquark state becomes negative. 

\section{Magnetic moments}

In this section, we study the magnetic moments of the lowest exotic 
pentaquark configuration with positive and negative parity. Which one 
of these is the ground state, depends on the relative size of the 
orbital excitations and the spin-flavor splittings. 

The magnetic moment is a crucial ingredient in calculations of the 
photo- and electroproduction cross sections of pentaquarks 
\cite{Nam,Zhao,photo}. A compilation of theoretical 
values for the chiral soliton model, different correlated quark models, 
the MIT bag model and for QCD sum rules has been presented in \cite{Liu}. 
To the best of our knowledge, the present calculation is the first one 
for an uncorrelated or constituent quark model. 

The magnetic moment of a multiquark system is given by the 
sum of the magnetic moments of its constituent parts
\ba
\vec{\mu} \;=\; \vec{\mu}_{\rm spin} + \vec{\mu}_{\rm orb} \;=\; 
\sum_i \mu_i (2\vec{s}_{i} + \vec{\ell}_{i}) ~, 
\ea
where $\mu_i=e_i/2m_i$, $e_i$ and $m_i$ represent the magnetic moment, 
the electric charge and the constituent mass of the $i$-th (anti)quark. 
The quark magnetic moments $\mu_u$, $\mu_d$ and $\mu_s$ are determined 
from the proton, neutron and $\Lambda$ magnetic moments to be $\mu_u=1.852$ 
$\mu_N$, $\mu_d=-0.972$ $\mu_N$ and $\mu_s=-0.613$ $\mu_N$ \cite{PDG}. 
The magnetic moments of the antiquarks satisfy $\mu_{\bar{q}}=-\mu_q$. 

\subsection{Negative parity pentaquark}

The lowest negative parity pentaquark state belongs to the 
$[f]_t=[42111]_{F_2}$ spin-flavor multiplet, is associated 
with the orbital state $L^p_t=0^+_{A_1}$ and has angular momentum 
and parity $J^p=1/2^-$ (see Fig.~\ref{orbex1}). 
The corresponding wave function is given by 
\ba
\psi_{A_2} &=& \left[ \psi^{\rm c}_{F_1} \times \left[ \psi^{\rm o}_{A_1} 
\times \psi^{\rm sf}_{F_2} \right]_{F_2} \right]_{A_2} ~,
\label{wfneg}
\ea
where the spin-flavor part can be expressed as a product of the 
antidecuplet flavor wave function $\phi_E$ 
and the $s=1/2$ spin wave function $\chi_{F_2}$ 
\ba
\psi^{\rm sf}_{F_2} &=& \left[ \phi_{E} \times \chi_{F_2} \right]_{F_2} ~.
\ea
Since the orbital wave function has $L^p_t=0^+_{A_1}$, the magnetic moment 
only depends on the spin part. For the $\Theta^+$, $\Xi_{3/2}^+$ and 
$\Xi_{3/2}^{--}$ exotic states we obtain
\ba
\mu_{\Theta^+} &=& \frac{1}{3}(2\mu_u + 2\mu_d + \mu_s) 
\;=\; 0.382 \; \mu_N ~,
\nonumber\\
\mu_{\Xi_{3/2}^{--}} &=& \frac{1}{3}(\mu_u + 2\mu_d + 2\mu_s) 
\;=\; -0.439 \; \mu_N ~,
\nonumber\\
\mu_{\Xi_{3/2}^+}    &=& \frac{1}{3}(2\mu_u + \mu_d + 2\mu_s)
\;=\; 0.502 \; \mu_N ~.
\label{mmneg}
\ea
These results are independent of the orbital wave functions, 
and are valid for any quark model in which the eigenstates have good 
$SU_{\rm sf}(6)$ spin-flavor symmetry. 
In the limit of equal quark masses $m_u=m_d=m_s$, the magnetic 
moments become proportional to the electric charge 
$\mu_{\Xi_{3/2}^{--}}= -2 \mu_{\Xi_{3/2}^+} = -2 \mu_{\Theta^+}$. 

\subsection{Positive parity pentaquark}

The lowest positive parity pentaquark has quantum numbers 
$[f]_t=[51111]_{A_1}$ with orbital excitation $L^p_t=1^-_{F_2}$ 
and angular momentum and parity $J^p=1/2^+$ (see Fig.~\ref{orbex2}). 
The wave function is now given by 
\ba
\psi_{A_2} \;=\; \left[ \psi^{\rm c}_{F_1} \times \left[ \psi^{\rm o}_{F_2} 
\times \psi^{\rm sf}_{A_1} \right]_{F_2} \right]_{A_2} ~,
\label{wfpos}
\ea
where the spin-flavor part can be expressed as a product of the 
antidecuplet flavor wave function $\phi_E$ 
and the $s=1/2$ spin wave function $\chi_E$ 
\ba
\psi^{\rm sf}_{A_1} &=& \left[ \phi_{E} \times \chi_{E} \right]_{A_1} ~. 
\ea
In this case, there are contributions to the magnetic moment from both 
the orbital angular momentum and the spin. Whereas the spin part does 
not depend on the orbital wave functions, the orbital part obviously 
does. As a result, the magnetic moments of the positive parity exotic 
pentaquarks $\Theta^+$, $\Xi_{3/2}^+$ and $\Xi_{3/2}^{--}$ are all equal 
\ba
\mu_{\Theta^+} \;=\; \mu_{\Xi_{3/2}^{--}} \;=\; \mu_{\Xi_{3/2}^+} 
\;=\; \frac{1}{3}(\mu_u + \mu_d + \mu_s) \;=\; 0.089 \; \mu_N ~.
\label{mmpos}
\ea
In this calculation we have used harmonic oscillator wave functions 
with $N=1$. In the limit of equal quark masses $m_u=m_d=m_s$, 
the moments vanish due to a cancellation between the spin and orbital 
contributions. 

\subsection{Results}

The results obtained for the magnetic moments are valid for any quark 
model in which the eigenstates have good $SU_{\rm sf}(6)$ spin-flavor 
symmetry. 
The magnetic moments for negative parity pentaquarks of Eq.~(\ref{mmneg}) 
are typically an order of magnitude smaller than the proton magnetic moment, 
whereas for positive parity they are even smaller due to a cancellation 
between orbital and spin contributions, see Eq.~(\ref{mmpos}). 
The magnetic moment of the $\Theta^+$ pentaquark is found to be 
0.382 $\mu_N$ for negative parity, and 0.089 $\mu_N$ for positive parity.

In Table~\ref{mmpenta} we present a comparison with the theoretical 
predictions of the magnetic moments of exotic pentaquarks 
for the chiral soliton model \cite{Kim1}, a diquark-diquark-antiquark 
bound state \cite{Nam,Zhao,Liu}, a diquark-triquark bound state \cite{Nam,Liu} 
the MIT bag model \cite{Zhao,Liu}, and for light cone QCD sum rules 
\cite{Huang}. Although the different models show some variations in the 
numerical values, generally speaking, the results for the $\Theta^+$ are 
relatively close, especially in comparison with the magnetic moment of the 
proton they are all small. 
For the $\Xi$ cascade pentaquarks there is a larger spread 
in the theoretical values which is mostly due to the result obtained for the 
$\Xi_{3/2}^{--}$ in the MIT bag model. The magnetic moments for 
positive parity pentaquarks tend to be smaller than those for negative parity. 

\section{Stringlike algebraic model}

After the schematic calculation of the previous section, we discuss now 
briefly the mass spectrum of exotic pentaquarks in a 
stringlike algebraic model. 
Hadronic spectra are characterized by the occurrence of linear Regge 
trajectories, i.e. $M^2 \sim \alpha L$, with almost identical slopes for 
baryons $\alpha_B=1.068$ (GeV)$^2$ \cite{BIL} and mesons $\alpha_M=1.092$ 
(GeV)$^2$ \cite{meson}. Such a behavior is also expected 
on basis of soft QCD strings in which the strings elongate as they rotate 
\cite{soft}. In the same spirit as in algebraic models of stringlike 
hadrons \cite{BIL,meson}, we use the mass-squared operator. The  
spin-flavor part is expressed in a G\"ursey-Radicati form \cite{GR}, 
i.e. in terms of Casimir invariants of the spin-flavor group chain of 
Eq.~(\ref{sfchain}) 
\ba
M^2 &=& M_0^2 + M^2_{\rm vib} + \alpha \, L 
   + a \, C_{2SU_{\rm sf}(6)} 
   + b \, C_{2SU_{\rm f}(3)} 
   + c \, C_{2SU_{\rm s}(2)} 
\nonumber\\
&& + d \, C_{1U_{\rm Y}(1)} 
   + e \, C_{1U_{\rm Y}(1)}^2  
   + f \, C_{2SU_{\rm I}(2)} ~.
\label{mass}
\ea
The coefficients $\alpha$, $a$, $b$, $c$, $d$, $e$ and $f$ are taken 
from a previous study of the nonstrange and strange baryon resonances 
\cite{BIL}. The constant $M_0^2$ is determined by identifying the 
ground state exotic pentaquark with the recently observed 
$\Theta^+(1540)$ resonance. Since the lowest orbital states with $L^p=0^+$ 
and $1^-$ are interpreted as rotational states, there is no 
contribution from the vibrational term $M^2_{\rm vib}$ to the mass of the 
corresponding pentaquark states.  

Just as for the harmonic oscillator case, the lowest pentaquark state is a 
flavor antidecuplet state with spin $s=1/2$ and isospin $I=0$, in 
agreement with the available experimental information which indicates that 
the $\Theta^+(1540)$ is an isosinglet \cite{saphir,clas,hermes}. 
In the present calculation, the ground state pentaquark has angular momentum 
and parity $J^p=1/2^-$, in agreement with recent work on QCD sum 
rules \cite{sumrule}, lattice QCD \cite{lattice} and a quark model 
calculation \cite{cqmneg}. 
It belongs to the $[42111]_{F_2}$ spin-flavor multiplet  
and the orbital excitation $0^+_{A_1}$. The first excited state at 1599 MeV 
is an isospin triplet with strangeness $S=+1$ of the 27-plet with the 
same value of angular momentum and parity $J^p=1/2^-$. The lowest 
pentaquark state with positive parity occurs at 1668 MeV. In the absence of 
a spin-orbit coupling, in this case we have a doublet with angular momentum 
and parity $J^p=1/2^+$, $3/2^+$.

The wave function of the ground state pentaquark has the same general 
structure as that of Eq.~(\ref{wfneg}) for the harmonic oscillator 
quark model 
\ba
\psi_{A_2} &=& \left[ \psi^{\rm c}_{F_1} \times \left[ \psi^{\rm o}_{A_1} 
\times \psi^{\rm sf}_{F_2} \right]_{F_2} \right]_{A_2} ~. 
\ea
They differ with respect to the treatment of the orbital part of the 
wave function, but the color-spin-flavor part is the same. Since the 
orbital wave function has $L^p_t=0^+_{A_1}$, the magnetic moments 
only depend on the spin part, and hence the magnetic moments in a 
stringlike algebraic model are the same 
as those for the harmonic oscillator quark model of Eq.~(\ref{mmneg}).

\section{Summary and conclusions}

The recent experimental evidence for the existence of exotic baryons has 
prompted an enormous amount of theoretical activity to interpret the data 
and to help understand its properties. The different approaches can 
be divided roughly into chiral soliton models which motivated the experiments, 
QCD related studies (large $N_{\rm c}$ limit, lattice calculations, 
sum rules), correlated quark models in which the exotic baryons arise  
as a quark cluster state, and uncorrelated or constituent quark models 
in which the exotic pentaquarks are interpreted as $q^4 \bar{q}$ bound 
states. 

In this article, we have analyzed the mass spectrum and the magnetic 
moments of exotic pentaquarks in a constituent quark model. First, we 
constructed a complete classification scheme based on two conditions: 
the pentaquark wave function is a color singlet, and is antisymmetric 
under any permutation of the four quarks. Exotic pentaquark states 
occur only in the flavor antidecuplet, the 27-plet and the 35-plet. 
The basis states in correlated quark models form a subset of the ones 
we have constructed. The precise ordering of pentaquark states in the 
mass spectrum depends on the choice of a specific dynamical model 
(Skyrme, chiral potential, Goldstone Boson Exchange, instanton, 
hypercentral, stringlike, ...). A schematic calculation in a harmonic 
oscillator quark model shows that the ground state configuration 
depends on the relative size of the orbital excitations and the 
spin-flavor splittings. 

For the calculation of the photo- and electroproduction cross sections 
of pentaquarks an important ingredient is the magnetic moment. Hereto we     
have derived the magnetic moments of the antidecuplet pentaquarks 
for both parities. For negative parity, they are typically an order of 
magnitude smaller than the proton magnetic moment, whereas for positive 
parity there is an additional reduction due to a cancellation between 
orbital and spin contributions. The magnetic moment of the $\Theta^+$ 
pentaquark is found to be 0.382 $\mu_N$ for negative parity, and 0.089 
$\mu_N$ for positive parity. 

Finally, we investigated the spectroscopy of exotic pentaquarks in an 
algebraic stringlike quark model which uses a generalized G\"ursey-Radicati 
form for the spin-flavor contributions. 
The interaction strenghts were taken from 
previous work on $q^3$ baryons. As a result, we obtained a ground state 
antidecuplet pentaquark with $J^p=1/2^-$ and isospin $I=0$, in agreement 
with experimental evidence that the $\Theta^+(1540)$ is an isosinglet.  
The first excited state at 1599 MeV is a $\Theta_1$ isospin triplet  
with strangeness $S=+1$ of the 27-plet with the 
same value of angular momentum and parity $J^p=1/2^-$. 
The antidecuplet state with strangeness $S=-2$ and isospin 
$I=3/2$ is calculated at an energy of 1956 MeV, to be compared with 
the recently observed $\Xi_{3/2}^{--}$ resonance at 1862 MeV \cite{cern}. 
The lowest pentaquark state with positive parity occurs at 1668 MeV. 
In the absence of a spin-orbit coupling, in this case we have a doublet 
with angular momentum and parity $J^p=1/2^+$, $3/2^+$. 
The magnetic moment of the $\Theta^+$ ground state pentaquark is 
0.382 $\mu_N$, the same value as for the harmonic oscillator quark model. 

In conclusion, the spectroscopy of exotic baryons will be a key testing 
ground for models of baryons and their structure. Especially the measurement 
of the angular momentum and parity of the $\Theta^+(1540)$ may help to 
distinguish between different models and to gain more insight into 
the relevant degrees of freedom and the underlying dynamics that determines 
the properties of exotic baryons. 

\section*{Acknowledgements}

This work is supported in part by a grant from CONACyT, M\'exico.

\clearpage

\begin{figure}
\centering
\Large
\setlength{\unitlength}{1pt}
\begin{picture}(180,200)(0,0)
\thicklines

\put(  0,  0) {\line(1,0){180}}
\put(  0,200) {\line(1,0){180}}
\put(  0,  0) {\line(0,1){200}}
\put(180,  0) {\line(0,1){200}}

\put( 70, 60) {\circle{16}}
\put(100, 50) {\circle{16}}
\put( 60,120) {\circle{16}}
\put( 90,150) {\circle{16}}
\put(120,100) {\circle{16}}

\put( 68,120) {\line( 1, 0){12}}
\put( 80,120) {\line( 1, 3){ 8}}
\put( 80,120) {\line( 2,-3){20}}
\put(100, 90) {\line( 2, 1){13}}
\put(100, 90) {\line(-1,-2){10}}
\put( 90, 70) {\line(-2,-1){13}}
\put( 90, 70) {\line( 1,-2){ 7}}

\put( 50, 50) {$q$}
\put(110, 30) {$q$}
\put( 40,120) {$q$}
\put( 96,168) {$q$}
\put(136,108) {$\bar{q}$}
\end{picture}
\caption[]{Pentaquark configuration $q^4 \bar{q}$}
\label{gsbar}
\end{figure}
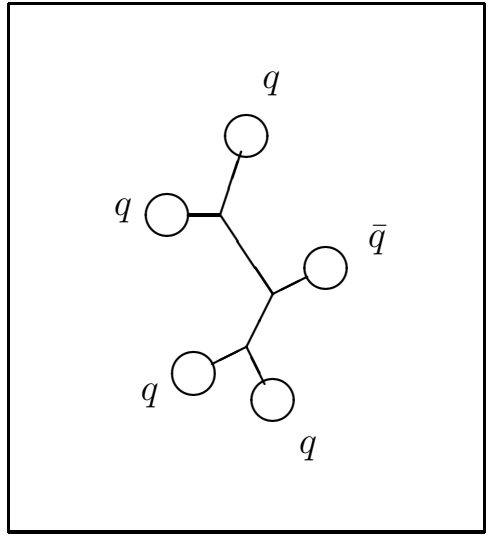

\clearpage

\begin{table}
\centering
\caption[]{Allowed spin, flavor and spin-flavor pentaquark states}
\vspace{15pt}
\label{states}
$\begin{array}{cccc}
\hline
& & & \\
& qqqq\bar{q} & \mbox{Dimension} & S_4 \sim {\cal T}_d \\
& & & \\
\hline
& & & \\
\mbox{spin} & [5] & 6 & A_1 \\
& [41] & 4 & A_1, F_2 \\
& [32] & 2 & F_2, E   \\
& & & \\
\hline
& & & \\
\mbox{flavor} & [51] & \mbox{35-plet} & A_1 \\
& [42]  & \mbox{27-plet} & F_2 \\
& [33]  & \mbox{antidecuplet} & E \\
& [411] & \mbox{decuplet} & A_1, F_2 \\
& [321] & \mbox{octet} & F_2, E, F_1 \\
& [222] & \mbox{singlet} & F_1 \\
& & & \\
\hline
& & & \\
\mbox{spin-flavor} 
& [51111]  &  700 & A_1 \\
& [411111] &   56 & A_1, F_2 \\
& [42111]  & 1134 & F_2 \\
& [321111] &   70 & F_2, E, F_1 \\
& [33111]  &  560 & E \\
& [32211]  &  540 & F_1 \\
& [222111] &   20 & F_1, A_2 \\
& [22221]  &   70 & A_2 \\
& & & \\
\hline
\end{array}$
\end{table}

\begin{table}
\centering
\caption[]{Discrete symmetry of exotic pentaquark states}
\vspace{15pt}
\label{tdpenta}
\begin{tabular}{cccccc}
\hline
& & & & & \\
$\psi$ & $\psi^{\rm c}$ & $\psi^{\rm osf}$ & $\psi^{\rm o}$ 
& $\psi^{\rm sf}$ & Exotic spin-flavor \\
& & & & & configuration \\
& & & & & \\
\hline
& & & & & \\
$A_2$ & $F_1$ & $F_2$ & $A_1$ & $F_2$ & $[42111]$ \\
& & & & & \\
$A_2$ & $F_1$ & $F_2$ & $F_2$ & $A_1$ & $[51111]$ \\
      & & & & $F_2$ & $[42111]$ \\
      & & & & $E$   & $[33111]$ \\
      & & & & $F_1$ & $[32211]$ \\
& & & & & \\
\hline
\end{tabular}
\end{table}

\begin{table}[ht]
\centering
\caption[]{Comparison of magnetic moments in $\mu_N$ of exotic 
antidecuplet pentaquarks with angular momentum $J=\frac{1}{2}$ for 
both positive and negative parity.}
\label{mmpenta}
\vspace{15pt}
\begin{tabular}{ccrrcrr}
\hline
& & & & & & \\
& \multicolumn{3}{c}{Positive parity} 
& \multicolumn{3}{c}{Negative parity}\\
& $\Theta^+$ & $\Xi_{3/2}^+$ & $\Xi_{3/2}^{--}$ 
& $\Theta^+$ & $\Xi_{3/2}^+$ & $\Xi_{3/2}^{--}$ \\
& & & & & & \\
\hline
& & & & & & \\
Present & 0.09 & 0.09 & 0.09 & 0.38 & 0.50 & --0.44 \\
& & & & & & \\
\protect{\cite{Kim1}} & 0.12 & 0.12 & --0.24 & & & \\
& 0.20 & 0.20 & --0.40 & & & \\
& 0.30 & 0.30 & --0.60 & & & \\
\protect{\cite{Nam}} & 0.18 & & & 0.49 & & \\
& 0.36 & & & 0.31 & & \\
\protect{\cite{Huang}}$^{\ast}$ & & & & $0.12 \pm 0.06$ & & \\
\protect{\cite{Liu,Zhao,JW}} & 0.08 & --0.06 &   0.12 & & & \\
\protect{\cite{Liu,SZ}} & 0.23 &   0.33 & --0.17 & & & \\
\protect{\cite{Liu,KL}} & 0.19 &   0.13 & --0.43 & & & \\
\protect{\cite{Liu,Zhao,DS}} & & & & 0.63 & 0.87 & --1.74 \\
& & & & & & \\
\hline
& & & & & & \\
\multicolumn{6}{l}{$^{\ast}$ Absolute value}
\end{tabular}
\end{table}


\begin{thebibliography}{99}

\bibitem{leps} 
LEPS Collaboration, T. Nakano et al., Phys. Rev. Lett. {\bf 91}, 012002 (2003). 

\bibitem{diana}
DIANA Collaboration, V.V. Barmin et al., Phys. Atom. Nucl. {\bf 66}, 1715 (2003). 

\bibitem{saphir}
SAPHIR Collaboration, J. Barth et al.,  
Phys. Lett. B {\bf 572}, 127 (2003). 

\bibitem{clas}
CLAS Collaboration, 
S. Stepanyan et al., Phys. Rev. Lett. {\bf 91}, 252001 (2003); \\
V. Kubarovsky and S. Stepanyan, hep-ex/0307088; \\
V. Kubarovsky et al., hep-ex/0311046; \\
H.G. Juengst, nucl-ex/0312019. 

\bibitem{itep}
A.E. Asratyan, A.G. Dolgolenko, and M.A. Kubantsev, 
hep-ex/0309042, to be published in Phys. Atom. Nucl. (2004).

\bibitem{hermes}
HERMES Collaboration, A. Airapetian et al., hep-ex/0312044. 

\bibitem{jlab}
See e.g. 
http://www.jlab.org/intralab/calendar/archive03/pentaquark/program.html

\bibitem{width}
R.A. Arndt, I.I. Strakovsky and R.L. Workman, 
Phys. Rev. C {\bf 68}, 042201 (2003);\\
S. Nussinov, hep-ph/0307357;\\
J. Haidenbauer and G. Krein, hep-ph/0309243;\\
R.N. Cahn and G.H. Trilling, hep-ph/0311245.

\bibitem{cern}
NA49 Collaboration, C. Alt et al., hep-ex/0310014.

\bibitem{soliton}
D. Diakonov, V. Petrov and M. Polyakov, 
Z. Phys. A {\bf 359}, 305 (1997);\\
H. Weigel, Eur. Phys. J. A {\bf 2}, 391 (1998);\\
M. Prasza{\l}owicz, in {\it Skyrmions and Anomalies}, 
Eds. M. Jezabek and M. Prasza{\l}owicz,
(World Scientific; Singapore, 1987); \\
M. Prasza{\l}owicz, Phys. Lett. B {\bf 575}, 234 (2003).

\bibitem{JW}
R. Jaffe and F. Wilczek, Phys. Rev. Lett. {\bf 91}, 232003 (2003).

\bibitem{SZ}
E. Shuryak and I. Zahed, hep-ph/0310270.

\bibitem{KL}
M. Karliner and H.J. Lipkin, Phys. Lett. B {\bf 575}, 249 (2003).

\bibitem{cqmpos}
Fl. Stancu, Phys. Rev. D {\bf 58}, 111501 (1998);\\
C. Helminen and D.O. Riska, 
Nucl. Phys. A {\bf 699}, 624 (2002);\\ 
A. Hosaka, Phys. Lett. B {\bf 571}, 55 (2003);\\
L.Ya. Glozman, 
Phys. Lett. B {\bf 575}, 18 (2003);\\
Fl. Stancu and D.O. Riska, Phys. Lett. B {\bf 575}, 242 (2003);\\
C.E. Carlson, Ch.D. Carone, H.J. Kwee and V. Nazaryan, 
Phys. Lett. B {\bf 579}, 52 (2004).

\bibitem{sumrule}
S.L. Zhu, Phys. Rev. Lett. {\bf 91}, 232002 (2003);\\
J. Sugiyama, T. Doi and M. Oka, hep-ph/0309271.

\bibitem{lattice}
F. Csikor, Z. Fodor, S.D. Katz and T.G. Kov\'acs, hep-lat/0309090;\\
S. Sasaki, hep-lat/0310014.

\bibitem{cqmneg}
C.E. Carlson, Ch.D. Carone, H.J. Kwee and V. Nazaryan,
Phys. Lett. B {\bf 573}, 101 (2003). 

\bibitem{parity}
T. Hyodo, A. Hosaka and E. Oset, 
Phys. Lett. B {\bf 579}, 290 (2004);\\
Y. Oh, H. Kim and S.H. Lee, hep-ph/0310019 and hep-ph/0312229;\\
W. Liu, C.M. Ko and V. Kubarovsky, 
hep-ph/0310087, Phys. Rev. C, in press;\\
B.-G. Yu, T.-K. Choi and C.-R. Ji, hep-ph/0312075;\\
A.W. Thomas, K. Hicks and A. Hosaka, hep-ph/0312083;\\
C. Hanhart, M. B\"uscher, W. Eyrich, K. Kilian, U.-G. Meissner, 
F. Rathmann, A. Sibirtsev and H. Str\"oher, hep-ph/0312236.

\bibitem{Nam}
S.I. Nam, A. Hosaka and H.-Ch. Kim, 
Phys. Lett. B {\bf 579}, 43 (2004).

\bibitem{Zhao}
Q. Zhao, hep-ph/0310350;\\
Q. Zhao and J.S. Al-Khalili, hep-ph/0312348.

\bibitem{photo}
K. Nakayama and K. Tsushima, hep-ph/0311112. 

\bibitem{BGS}
R. Bijker, M.M. Giannini and E. Santopinto, 
preprint hep-ph/0310281.

\bibitem{KM}
P. Kramer and M. Moshinsky, 
Nucl. Phys. {\bf 82}, 241 (1966);\\
Fl. Stancu, Phys. Rev. D {\bf 58}, 111501 (1998). 

\bibitem{GMO}
See e.g. M. Gell-Mann and Y. Ne'eman, 
{\it The eightfold way} (W.A. Benjamin, Inc., New York, 1964). 

\bibitem{GR}
F. G\"ursey and L.A. Radicati ,
Phys. Rev. Lett. {\bf 13}, 173 (1964).

\bibitem{BIL}
R. Bijker, F. Iachello and A. Leviatan,
Ann. Phys. (N.Y.) {\bf 236}, 69 (1994);\\
R. Bijker, F. Iachello and A. Leviatan,
Ann. Phys. (N.Y.) {\bf 284}, 89 (2000).

\bibitem{Riska}
L.Ya. Glozman and D.O. Riska, 
Phys. Rep. {\bf 268}, 263 (1996).

\bibitem{Liu}
Y.-R. Liu, P.-Z. Huang, W.-Z. Deng, X.-L. Chen and S.-L. Zhu, 
hep-ph/0312074.

\bibitem{PDG}
Particle Data Group, Phys. Rev. D {\bf 66}, 010001 (2002).

\bibitem{Kim1}
H.-C. Kim and M. Prasza{\l}owicz, hep-ph/0308242.

\bibitem{Huang}
P.-Z. Huang, W.-Z. Deng, X.-L. Chen and S.-L. Zhu, hep-ph/0311108.

\bibitem{DS}
D. Strottman, Phys. Rev. D {\bf 20}, 748 (1979). 

\bibitem{meson}
F. Iachello, N.C. Mukhopadhyay and L. Zhang,
Phys. Lett. B {\bf 256}, 295 (1991);\\
F. Iachello, N.C. Mukhopadhyay and L. Zhang,
Phys. Rev. D {\bf 44}, 898 (1991).
 
\bibitem{soft}
K. Johnson and C.B. Thorn,
Phys. Rev. D {\bf 13}, 1934 (1974);\\
I. Bars and A.J. Hanson,
Phys. Rev.D {\bf 13}, 1744 (1974).

\end{thebibliography}
\end{document}